\documentclass[
aip,
 amsmath,amssymb,
reprint,%
]{revtex4-1}

\usepackage{graphicx}
\usepackage{epstopdf}

\usepackage{dcolumn}
\usepackage{bm}

\usepackage[mathlines]{lineno}

\usepackage[utf8]{inputenc}
\usepackage[T1]{fontenc}
\usepackage{mathptmx}

\begin{document}

\preprint{AIP/123-QED}

\title{Radio-frequency stress-induced modulation of CdTe/ZnTe quantum dots}

\author{V. Tiwari}
\affiliation{Institut N\'{e}el, CNRS, Univ. Grenoble Alpes and Grenoble INP, 38000 Grenoble, France}

\author{K. Makita}
\affiliation{Institute of Materials Science, University of Tsukuba, Tsukuba 305-8573, Japan}

\author{M. Arino}
\affiliation{Institute of Materials Science, University of Tsukuba, Tsukuba 305-8573, Japan}

\author{M. Morita}
\affiliation{Institute of Materials Science, University of Tsukuba, Tsukuba 305-8573, Japan}

\author{T. Crozes}
\affiliation{Institut N\'{e}el, CNRS, Univ. Grenoble Alpes and Grenoble INP, 38000 Grenoble, France}

\author{E. Bellet-Amalric}
\affiliation{IRIG-PHELIQS, CEA and Univ. Grenoble Alpes, 38000 Grenoble, France}

\author{S. Kuroda}
\affiliation{Institute of Materials Science, University of Tsukuba, Tsukuba 305-8573, Japan}

\author{H. Boukari}
\affiliation{Institut N\'{e}el, CNRS, Univ. Grenoble Alpes and Grenoble INP, 38000 Grenoble, France}

\author{L. Besombes}\email{lucien.besombes@neel.cnrs.fr}
\affiliation{Institut N\'{e}el, CNRS, Univ. Grenoble Alpes and Grenoble INP, 38000 Grenoble, France}
\date{\today}

\begin{abstract}

We demonstrate radio-frequency tuning of the energy of individual CdTe/ZnTe quantum dots (QDs) by Surface Acoustic Waves (SAWs). Despite the very weak piezoelectric coefficient of ZnTe, SAW in the GHz range can be launched on a ZnTe surface using interdigitated transducers deposited on a c-axis oriented ZnO layer grown on ZnTe containing CdTe QDs. The photoluminescence (PL) of individual QDs is used as a nanometer-scale sensor of the acoustic strain field. The energy of QDs is modulated by SAW in the GHz range and leads to characteristic broadening of time-integrated PL spectra. The dynamic modulation of the QD PL energy can also be detected in the time domain using phase-locked time domain spectroscopy. This technique is in particular used for monitoring complex local acoustic fields resulting from the superposition of two  or more SAW pulses in a cavity. Under magnetic field, the dynamic spectral tuning of a single QD by SAW can be used to generate single photons with alternating circular polarization controlled in the GHz range.

\end{abstract}

\maketitle

\section{\label{sec1}Introduction}

Surface Acoustic Waves (SAWs), phonon like excitations bound to the surface of a solid, are attracting a lot of interest in quantum technologies \cite{Schuetz2015, Bienfait2019,Lemonde2018}. As SAWs couple to almost any quantum system and experience only weak dissipation in crystalline solids, they are proposed as quantum bus enabling long-range coupling of a wide range of $qubit$ including individual localized spins \cite{Golter2016}. SAW provide an effective classical field to control motional or internal states of a quantum system \cite{Whiteley2019} and could also be operated in the quantum regime at low temperature \cite{Satzinger2018,Moores2018}.

II-VI semiconductors can be doped with transition metal magnetic elements offering a large choice of localized spins. Individual magnetic atoms can be optically probed and to some extent controlled when they are incorporated in a quantum dot (QD)\cite{Besombes2004,Kudelski2007,Goryca2009,LeGall2011,Koenrad2011,Smolenski2016}. Among these magnetic elements, atoms featuring both spin and orbital degrees of freedom present a large spin-to-strain coupling. Such atoms could be used in hybrid spin-mechanical devices \cite{Lee2017,Rabl2010,Ovar2014,Tessier2014}. This is in particular the case of chromium incorporated as a Cr$^{2+}$ ion in CdTe/ZnTe QDs which carries both an electronic spin S=2 and an orbital momentum L=2 \cite{Lafuente2018}. A spin-to-strain coupling more than two orders of magnitude larger than that for elements without orbital momentum [ vacancy (NV) centers in diamond \cite{Tessier2014}, Mn atoms in II-VI semiconductors \cite{Lafuente2015}] was evidenced for Cr \cite{Lafuente2016,Vallin1974}.

It has been demonstrated that the spin states S$_z=\pm$1 of a Cr atom in CdTe can be prepared and readout optically and present a relaxation time larger than 10 $\mu s$ at low temperature \cite{Tiwari2020}. In analogy with the spin structure of NV centers in diamond, the spin states S$_z=\pm$1 of Cr in a self-assembled QD form a $qubit$ ($\lbrace+1;-1\rbrace$ spin $qubit$) strongly coupled to in-plane strain \cite{Lee2017}. SAWs, which generate in-plane strain along their direction of propagation, could therefore be used to coherently manipulate the S$_z=\pm1$ spin states of an individual Cr atom or for phonon assisted long range spin-spin coupling \cite{Lukin2013}. For such applications, SAWs in the GHz range are required to (i) overcome the eventual energy splitting of the S$_z=\pm1$ Cr spin states induced by a residual static in-plane strain in the CdTe/ZnTe QDs and (ii) to permit a mechanical coherent control of the $qubit$ in a few tens of nanoseconds.

SAWs are usually generated by a Radio Frequency (RF) signal with electro-mechanical transducers based on the piezoelectric effect \cite{Thevenard2019}. However, ZnTe grown on a (001) surface presents a very weak piezoelectric coefficient, about 10 times smaller than GaAs which is already a bad piezoelectric material \cite{Adachi2005}. To our knowledge SAW generation and propagation in ZnTe or similar zinc-blende II-VI materials has never been reported.

We show here that SAW in the GHz range can be launched on a ZnTe surface using interdigitated transducers (IDT) deposited on a c-axis oriented ZnO layer grown on ZnTe containing CdTe QDs. The photoluminescence (PL) of individual positively charged QDs is used as a nano-scale sensor of the acoustic strain field. The energy of QDs is modulated at high frequency by the strain field of the SAW resulting in characteristic broadening of time-integrated PL spectra. Under a longitudinal magnetic field, the dynamic spectral tuning of a single QD emission can be used to create high frequency modulated single photon sources with alternating circular polarization.

The dynamic modulation of the QD emission can also be detected in the time domain using phase-locked time domain spectroscopy. This technique is in particular used for monitoring complex local acoustic strain fields resulting from the superposition of two or more SAW pulses in a cavity. The excitation of a single IDT by short RF electrical pulses can generate mechanical pulses in the 100 nanoseconds range that can be reflected by a second IDT. Counter-propagating SAWs produced by two identical IDTs induce standing waves that can be controlled by the relative phase of the two SAWs. These interferences of acoustic strain fields are efficiently detected in the emission energy of a QD. The obtained high frequency and large in-plane strain component of the SAW in the 10$^{-4}$ range could be used for the coherent mechanical driving of the spin state of an individual magnetic atom.

\section{\label{sec2}SAW transducers on ZnTe containing CdTe quantum dots.}

SAW devices have never been realized in ZnTe or in any other tellurium or selenium based zinc-blende semiconductors. The generation of SAW from an electrical RF source requires an electro-mechanical transducer based on the piezoelectric effect. These II-VI materials grown on a (001) surface presents a very weak piezoelectric coefficient $e_{14}$. Furthermore, the relativity slow speed of phonons in these II-VI compounds compared to common III-V materials requires the realisation of smaller devices to reach the same electro-mechanical resonance frequency.

\begin{figure}[hbt]
\centering
\includegraphics[width=3.3in]{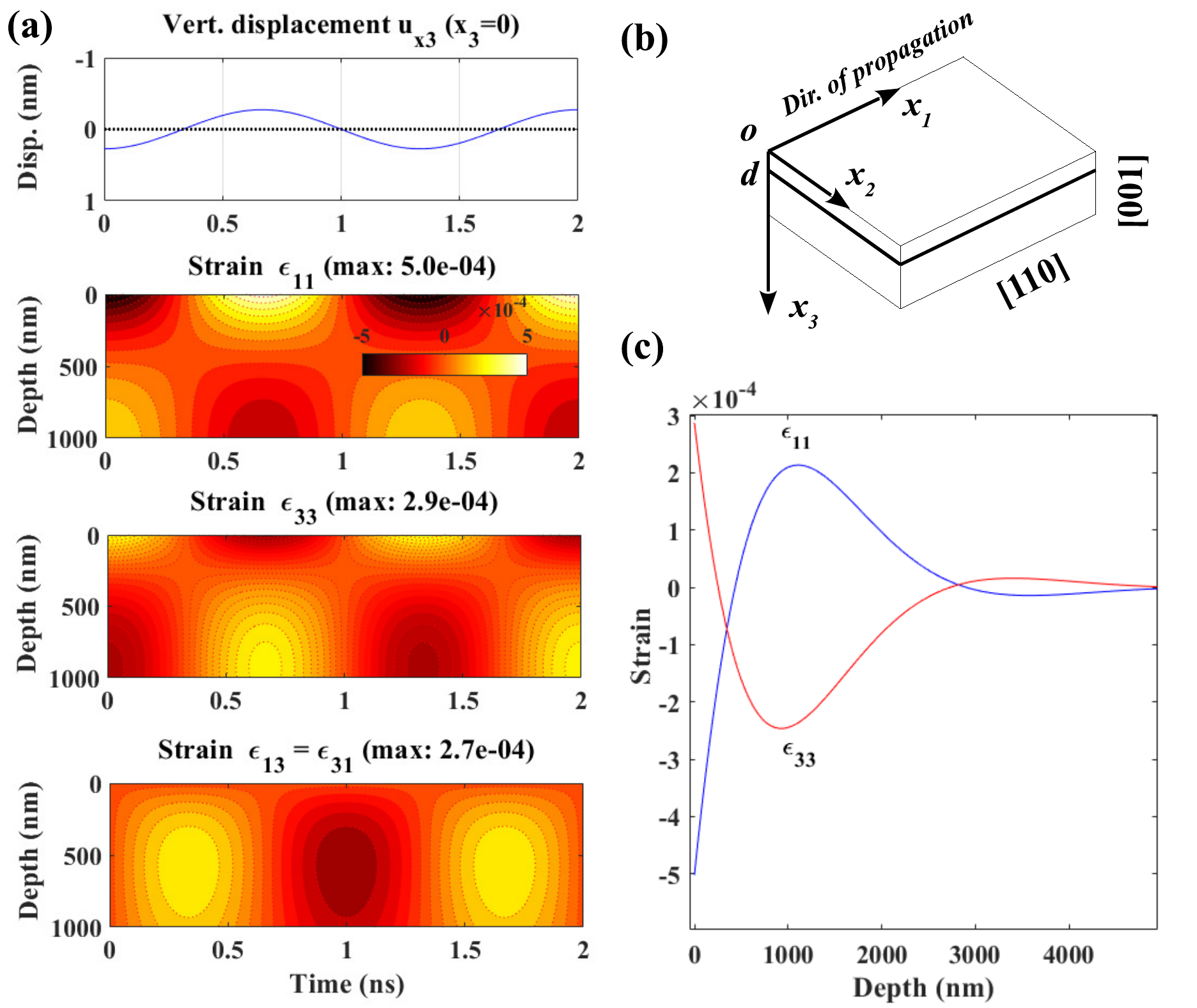}
\caption{Calculated strain field of a SAW at 0.75 GHz propagating along [110] in ZnTe. (a) Map of the time dependence of the profile along x$_3$ of the different strain components calculated for a maximum displacement of 0.27 nm of the surface (adjustable parameter A=0.2 nm, see appendix A). (b) Geometry of the device. (c) x$_3$ profile of $\epsilon_{11}$ and $\epsilon_{33}$ at a fixed time (t=0 ns).}
\label{Fig1}
\end{figure}

For the design of the devices, we used an analytical model to obtain the main parameters of SAW propagating along the [110] direction of ZnTe (see Appendix A). Such SAW generates strain along [110], the direction along which the largest dynamical spin to strain coupling is expected for a Cr atom \cite{Vallin1974}. Using the mechanical parameters of ZnTe (table 1 in Appendix A) we obtain, at low temperature, a speed of SAW along [110] $v_{S}\approx$ 1980 $m.s^{-1}$. This speed which controls the SAW wavelength $\lambda_{S}=v_{S}/f_{S}$ imposes the geometry of the transducer required to reach a given SAW frequency $f_{S}$. To generate SAWs inter-digitated electrodes (IDT) are usually used to electrically create dynamic periodical strain at the surface of the material. When $\lambda_{S}$ matches the period of the IDT, mechanical constructive interferences occur and SAW can be launched at the surface of the crystal. 

The calculated strain distribution for a SAW at 0.75 GHz propagating along [110] is presented in Fig. \ref{Fig1}. One should note that the strain is only present in the plane containing the normal of the surface and the direction of propagation. The strain field decays with the distance from the surface on a length scale of approximately $\lambda_{S}$. As expected, the highest strain field is obtained close to the surface.

The studied samples consist of a layer of self-assembled CdTe QDs grown by molecular beam epitaxy on a p-doped ZnTe (001) substrate \cite{Wojnar2011}. In the SAW devices the QD layer is placed at d=100 nm below the ZnTe surface. Thus, the QDs are close enough to the surface so that they can experience a large strain field but sufficiently far away from charge fluctuating surface states to limit the broadening of the PL lines.

\begin{figure}[hbt]
\centering
\includegraphics[width=3.3in]{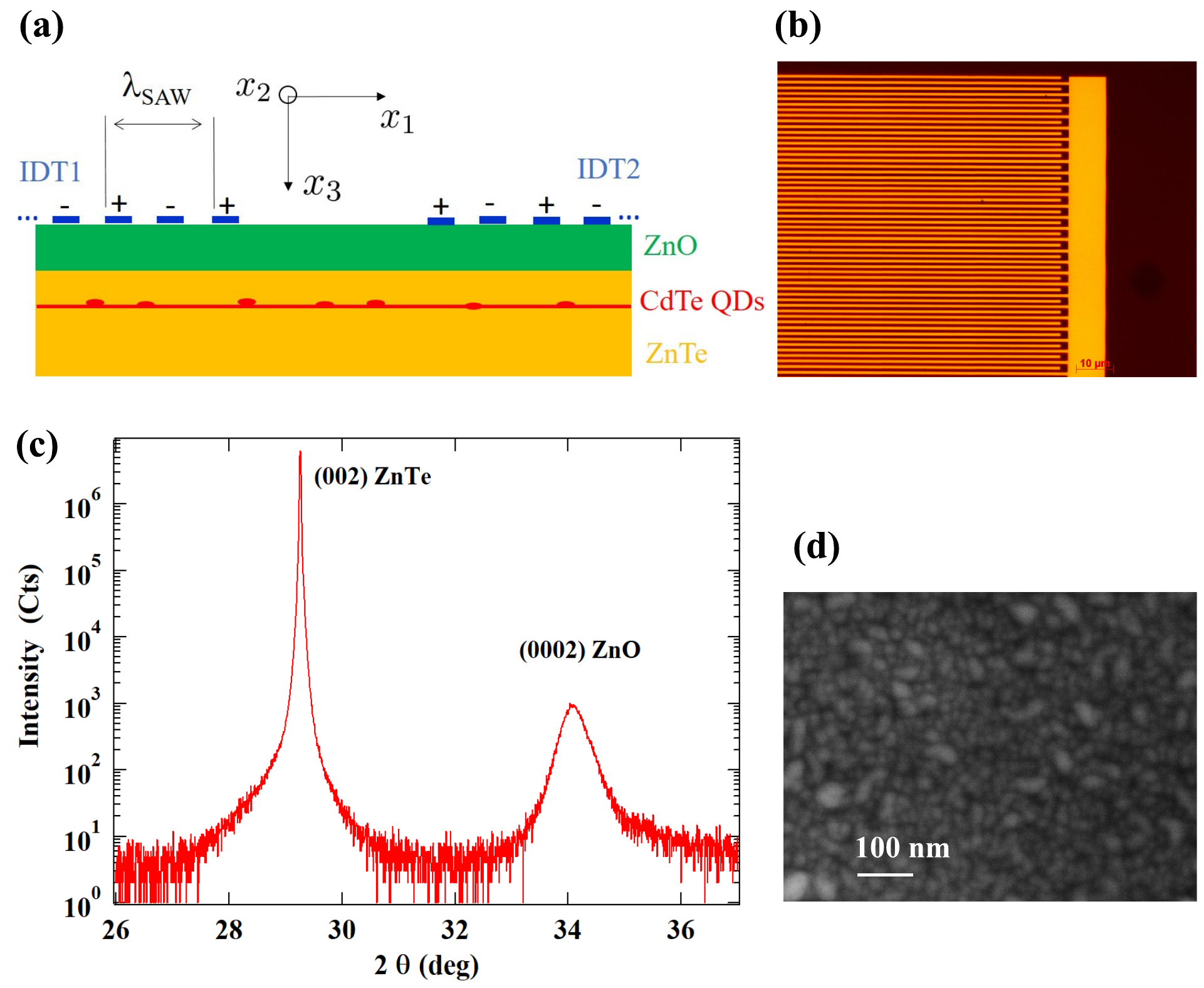}
\caption{(a) Scheme of a SAW device on a CdTe/ZnTe quantum dot structure. (b) Optical image of an aluminium IDT. (c) X ray characterization of the ZnO/ZnTe heterostructure: 2$\theta$-$\theta$ scan around the (002) ZnTe and (0002) ZnO reflections. (d) Scanning electron microscopy image of the ZnO surface.}
\label{Fig2}
\end{figure}

To electrically generate SAW we introduced a piezoelectric ZnO layer between the ZnTe containing the CdTe QDs and the IDT. The 180 nm thick ZnO is deposited by RF sputtering at a growth rate of about 110 nm per hour. An Argon/ Oxygen plasma with 5$\%$ of oxygen is used to limit the formation of oxygen vacancies that are known to induce an n-type doping of ZnO. An Ar plasma etching of the ZnTe surface is performed before the ZnO deposition to promote a good adhesion of the ZnO on the ZnTe. The ZnO is deposited at a moderate temperature of 200$^{\circ}$C to avoid any thermally driven intermixing of Cd and Zn at the interfaces of the CdTe/ZnTe QDs. 

A significant defect density is likely in these heterojunctions due to the differing symmetry of the hexagonal ZnO crystal and zinc blende ZnTe. Furthermore, the crystalline relation (0001)ZnO//(001)ZnTe and [1010]ZnO/[110]ZnTe results in a lattice mismatch of about 25$\%$. The crystalline structure of the ZnO layer was analysed by X-Ray diffraction (Fig. \ref{Fig2}(c)). A ZnO (0002) reflection is observed indicating the wurtzite c-plane ZnO (0001) orientation. The ZnO layer exhibits a columnar structure: the growth direction corresponds to the c axis but with a mosaic spread around 10 degrees and no in-plane epitaxial relation respect to the ZnTe substrate is observed in the X-Ray diffraction analysis. The columnar structure of the layer also appears in scanning electron microscopy images (Fig. \ref{Fig2}(d)) where grains are observed with a typical lateral size around 30 nm. 

The piezoelectric tensor of a 6mm material like ZnO is invariant by rotation around the c axis (Ox$_3$ axis). Mosaicity observed in the plane of the deposited ZnO layer is consequently not a problem for the efficiency of a SAW IDT. The important parameter for an efficient electro-mechanical conversion is the alignment of the c axis along Ox$_3$ perpendicular to the sample surface. The columnar structure has however some effects on the mechanical properties of the layer and can affect the SAW propagation by introducing some dissipation. In our case the ZnO layer is thin compared to the SAW wavelength and most of the mechanical wave propagates in ZnTe. The SAW characteristics are then expected to be mainly controlled by the parameters of ZnTe and the generated strain field close to the calculations presented in Fig.~\ref{Fig1}.

IDT formed by 30 pairs of fingers (N=60 fingers) with a pitch of w=750 nm (period 4w=3$\mu m$) were realized on the ZnO surface by e-beam lithography and lift-off of a 75 nm thick aluminium layer. With the calculated SAW speed in ZnTe this geometry should result in an electro-mechanical resonance at $f_0=v_{S}/4w\approx$0.67 GHz.

\section{\label{sec3}Acousto-optic spectral control of single charged quantum dots.}

Individual QDs in SAW devices were studied by optical micro-spectroscopy at liquid helium temperature (T=4.2 K). The PL of QDs was excited with a continuous wave (cw) dye laser tuned to an excited state, dispersed and filtered by a 1 $m$ double spectrometer before being detected by a Si cooled multichannel charged coupled device (CCD) camera or, for time resolved experiments, a Si avalanche photodiode (APD) with a time resolution of about 50 picoseconds. The laser power was stabilized by an electro-optic variable attenuator. Piezoelectric actuators and scanners were used to move the SAW device connected to coaxial RF cables in front of a high numerical aperture microscope objective (NA=0.85). A power stabilized and frequency tunable RF source permit to excite the IDTs. Internal modulation of the source was used to generate RF pulses in the 100 ns range. A magnetic field, up to 9 T along the growth axis of the QDs and 2 T in the plane of the dots, could be applied with a vectorial superconducting coil. 

\begin{figure}[hbt]
\centering
\includegraphics[width=3.3in]{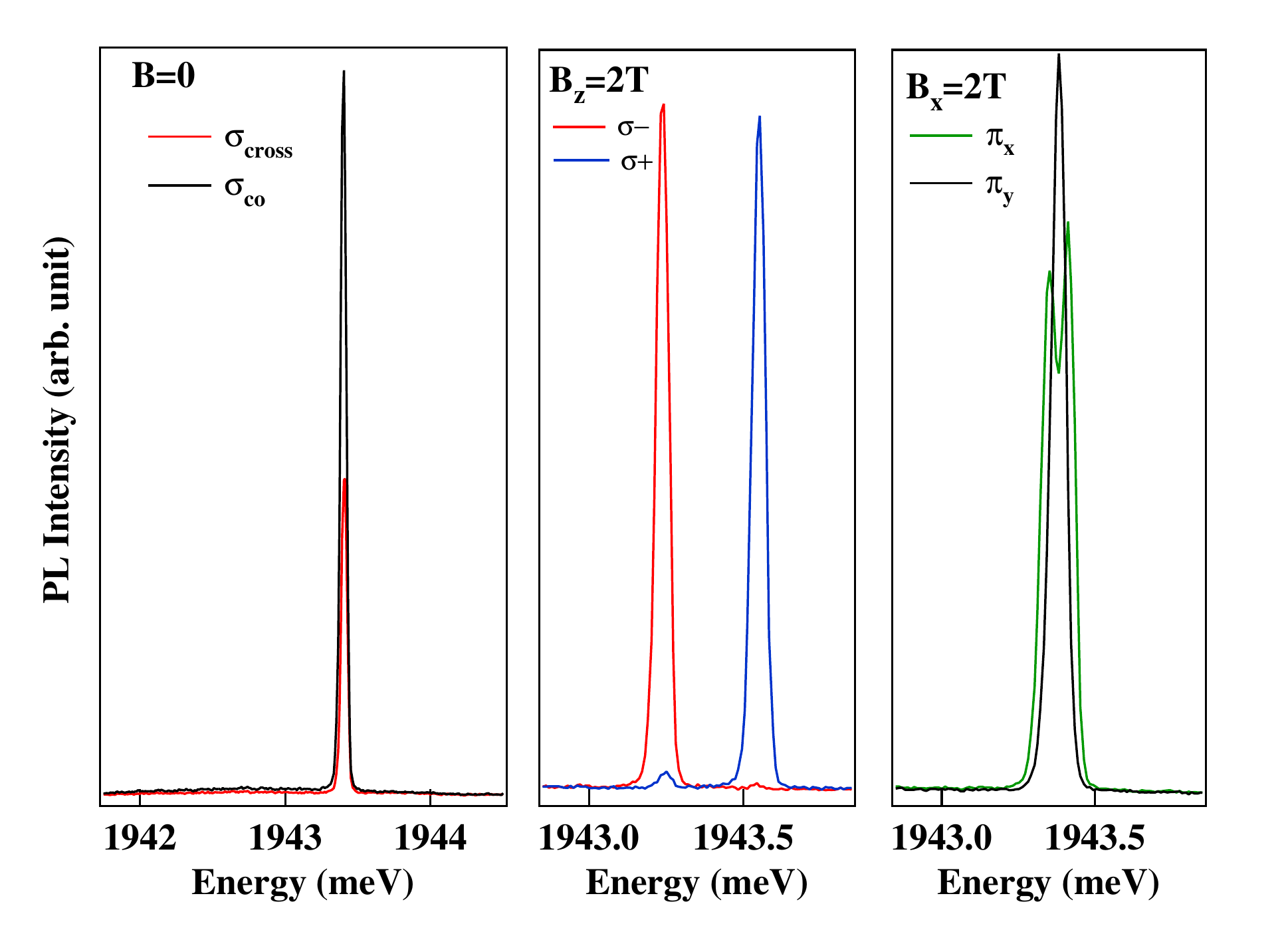}
\caption{Characteristic magnetic field dependence of the PL of a positively charged exciton in a CdTe/ZnTe QD (QD1) in a SAW device.}
\label{Fig2bis}
\end{figure}

As the QD size is much smaller than the wavelength of the SAW, $\lambda_{S}$, the strain field can be considered to be constant over the size of the QDs and an individual dot can be seen as a local probe of the strain field at the nanometer-scale. The strain field of the SAW is probed through the strain induced energy shift of the PL peak of positively charged QDs \cite{Poizat}. As charged excitons in QDs are not split by the electron-hole exchange interaction they present at zero magnetic field a single and narrow emission line. Such charged QDs can be identified by their polarization properties and their evolution under magnetic field (Fig. \ref{Fig2bis}). At zero field under circularly polarized excitation, the PL of positively charged exciton is co-polarized. Under a longitudinal magnetic field the emission is split by the Zeeman energy and perfectly circularly polarized whereas under a transverse field the PL becomes linearly polarized along and perpendicular to the applied field direction (Fig. \ref{Fig2bis}). 

\begin{figure}[hbt]
\centering
\includegraphics[width=3.3in]{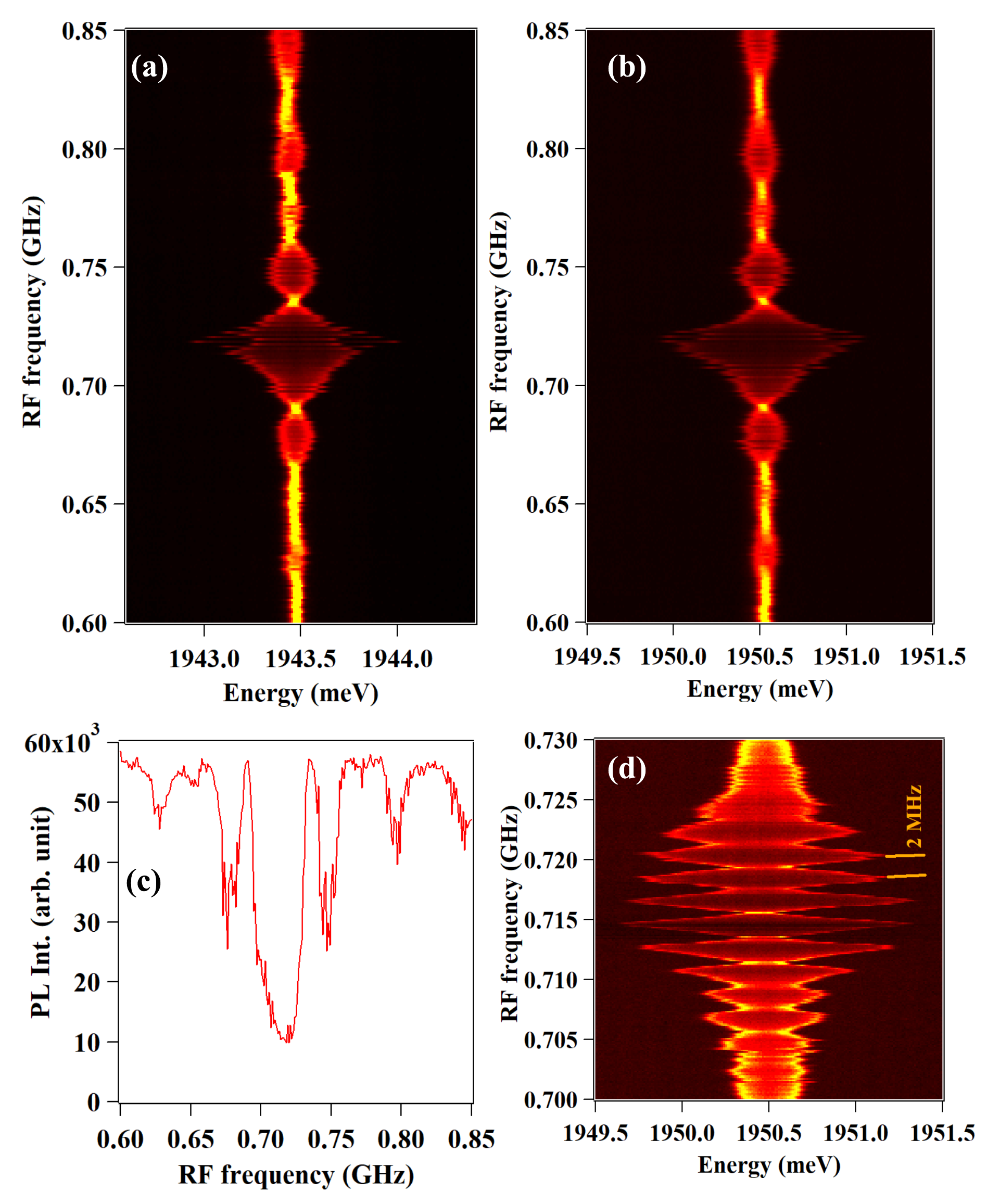}
\caption{RF frequency dependence of the emission of a charged exciton in two different QDs: (a) QD1 at 20 dBm and (b) QD2 at 21 dBm. (c) RF frequency dependence of the PL intensity of QD2 recorded around the central line showing the position of the main electro-mechanical resonance. (d) Detail of the mechanical resonance on QD2 showing a frequency modulation at 2 MHz arising from a cavity effect between two IDTs.}
\label{Fig3}
\end{figure}

To probe the coupling of the SAW to the QDs, the optical response of charged excitons is first detected with an integration time in the one second range. Figure \ref{Fig3} presents the spectral tuning of a single charged QD by the dynamic strain field of a SAW at variable frequencies. When the frequency of the RF excitation corresponds to the electro-mechanical resonance of the IDT, a large broadening is observed in the QD PL. In this configuration, the strain wave of SAW induces a sinusoidal modulation of the energy levels and PL lines oscillates around their equilibrium position. In time-integrated spectra, the dynamic energy modulation is averaged over many acoustic cycles. Emission lines spend more time at the minima and maxima of their oscillating energy and for sufficiently high RF power, peaks are observed at these points in the spectra. The overall width of the PL broadening maps the amplitude of the strain field.

The main mechanical resonance is observed around $f_0$=0.72 GHz (Fig. \ref{Fig3}(c)). This resonance frequency corresponds to a measured speed of SAW $v_{S}=4wf_0\approx$ 2160 m s$^{-1}$ slightly larger than the calculated value ($v_{S}\approx$ 1980 m s$^{-1}$). This small difference is likely due to partial propagation of the SAW in the thin ZnO layer where phonons propagate at a higher velocity. The full width at half maximum of the resonance, around 30 MHz, is close to the expected value given by $\Delta f\approx\frac{1.77}{N-1}f_0\approx$22 MHz \cite{BookOndesAcoustiques}. 

\begin{figure}[hbt]
\centering
\includegraphics[width=3.3in]{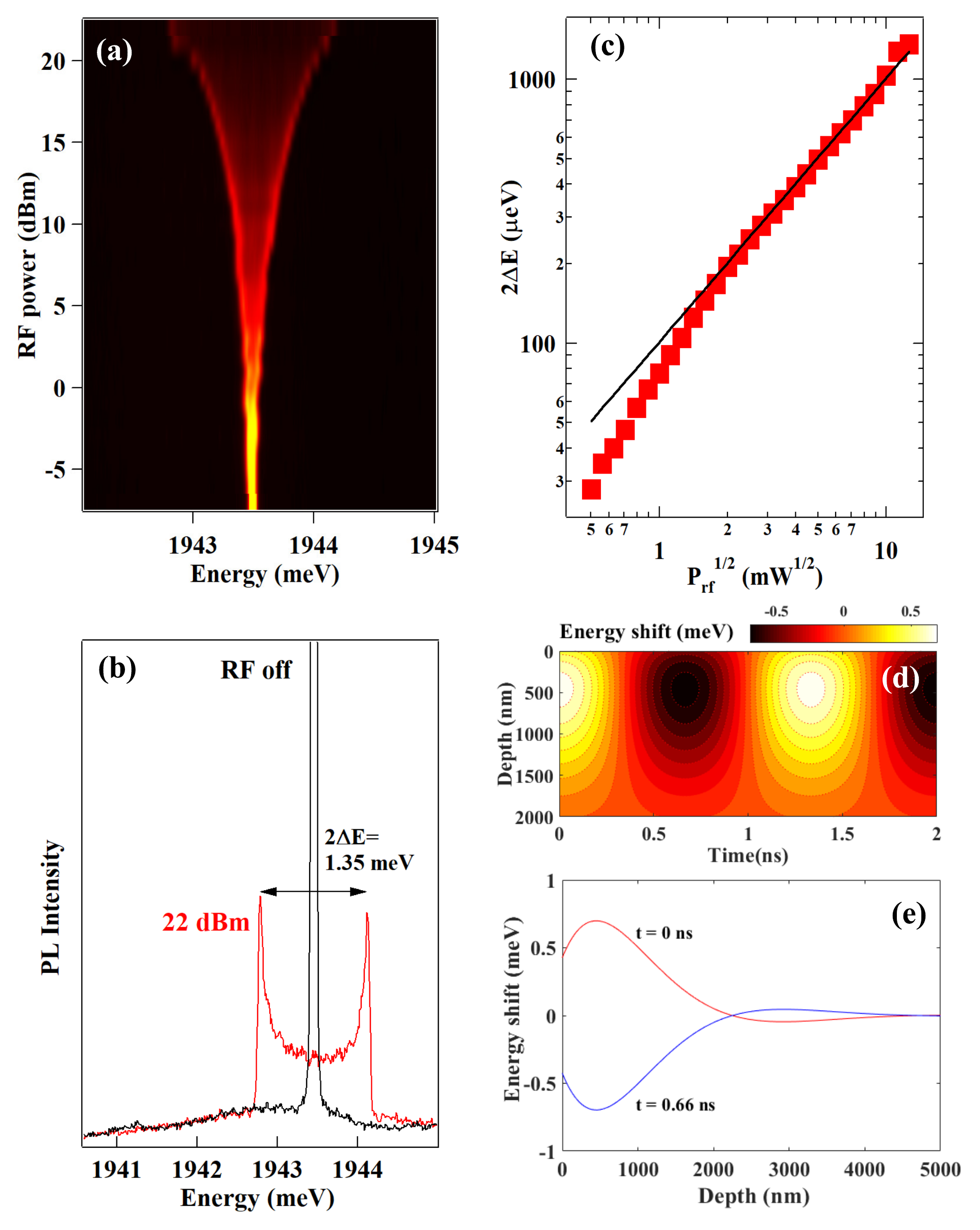}
\caption{(a) Intensity map of the RF power dependence of the SAW induced energy shift recorded on QD1 for an excitation at 0.718 GHz. (b) Comparison of PL obtained in the absence of RF excitation and at high RF excitation power (22 dBm). (c) Strain induced energy splitting as a function of the RF amplitude. The black line is a linear fit. (d) Map of the time dependence of the strained induced energy shift of a QD emission calculated for a SAW at 0.75 GHz and a maximum displacement of 0.27 nm of the surface (adjustable parameter A=0.2 nm, see appendix A). (e) Corresponding calculated depth profile of the energy shift.}
\label{Fig4}
\end{figure}

Within the mechanical resonance, a modulation with a period $\delta f \approx$ 2 MHz is observed in the frequency dependence of the energy shift (see Fig. \ref{Fig3}(d)). This modulation is due to a cavity effect induced by the reflection of the SAW launched by the IDT1 on the IDT2 that is not connected to an external circuit (see Fig. \ref{Fig2}). Note that an IDT left without electrical connections behaves like a Bragg reflector for SAWs \cite{BookOndesAcoustiques}. The two identical IDTs separated by $l=500 \mu m$ create an acoustic Fabry-Perrot cavity with a mode spacing $\delta f=v_{S}/2L_{cav}^{eff}$. From the measured mode spacing one can deduce an effective acoustic cavity length $L_{cav}^{eff}=\lambda_{S}f_{S}/2 \delta f\approx  540 \mu m$. This effective cavity length, slightly larger than the distance between the IDT, results from a partial penetration of the SAW inside the Bragg reflector. As it will be discussed in detail in section \ref{sec5}, the reflection of SAW can also be observed directly in the time domain under pulsed RF excitation.

At a fixed RF frequency, the strain induced broadening of the PL line is controlled by the RF excitation power. Figure \ref{Fig4}(a) shows the PL map of a QD modulated by SAW as a function of the energy and of the power of the RF excitation $P_{rf}$. The measured broadening labelled 2$\Delta E$ is extracted from the data and plotted in Fig. \ref{Fig4}(c) as a function $\sqrt(P_{rf})$. 2$\Delta E$ is linear with the amplitude of the strain field proportional to $\sqrt(P_{rf})$. Only a slight deviation from the expected linear dependence is observed at low excitation intensity. This deviation is likely to be induced by a depletion layer at the surface of the residually doped ZnO layer or to the presence of interface states at the Al/ZnO contact which can both reduce the applied electric field at small bias voltages.

In a CdTe/ZnTe QD the shift of the optical transition energy from the conduction to the heavy-hole band induced by the strain field of a SAW propagating along the [110] direction is given by (see appendix A):

\begin{eqnarray}
\Delta E_{g,hh}=(a_c-a_v)(\epsilon_{11}+\epsilon_{33})-\frac{b}{2}(\epsilon_{11}-2\epsilon_{33})
\end{eqnarray}

\noindent where $a_c=-3.96eV$, $a_v=0.55eV$ and $b=-1.0eV$ are the deformation potential of CdTe \cite{Adachi2005}. The corresponding calculated energy shift induced by a SAW at 0.75 GHz is presented in Fig. \ref{Fig4}(d). We found that for a maximum displacement of the atoms at the surface of $\pm0.27nm$, a maximum energy shift of $2\Delta E\approx1.4 meV$ is expected for QD located around 400 nm below the surface of the sample (Fig. \ref{Fig4}(e)). This energy shift depends on the depth and is linear with the maximum surface displacement. 

The measured energy shift $2\Delta E\approx1.35 meV$ at high RF excitation power (22 dBm in Fig. \ref{Fig4}(b)) shows that a strain of a few 10$^{-4}$ can be applied on the QDs layer (see strain field maps in Fig. \ref{Fig1}(a)). This modelling also suggests that simple non-resonant PL of a QD could be used to measure surface displacement in the order of 10 pm corresponding to an energy shift $2\Delta E=0.07 meV$, larger than the linewidth of the PL line. 

\begin{figure}[hbt]
\centering
\includegraphics[width=3.3in]{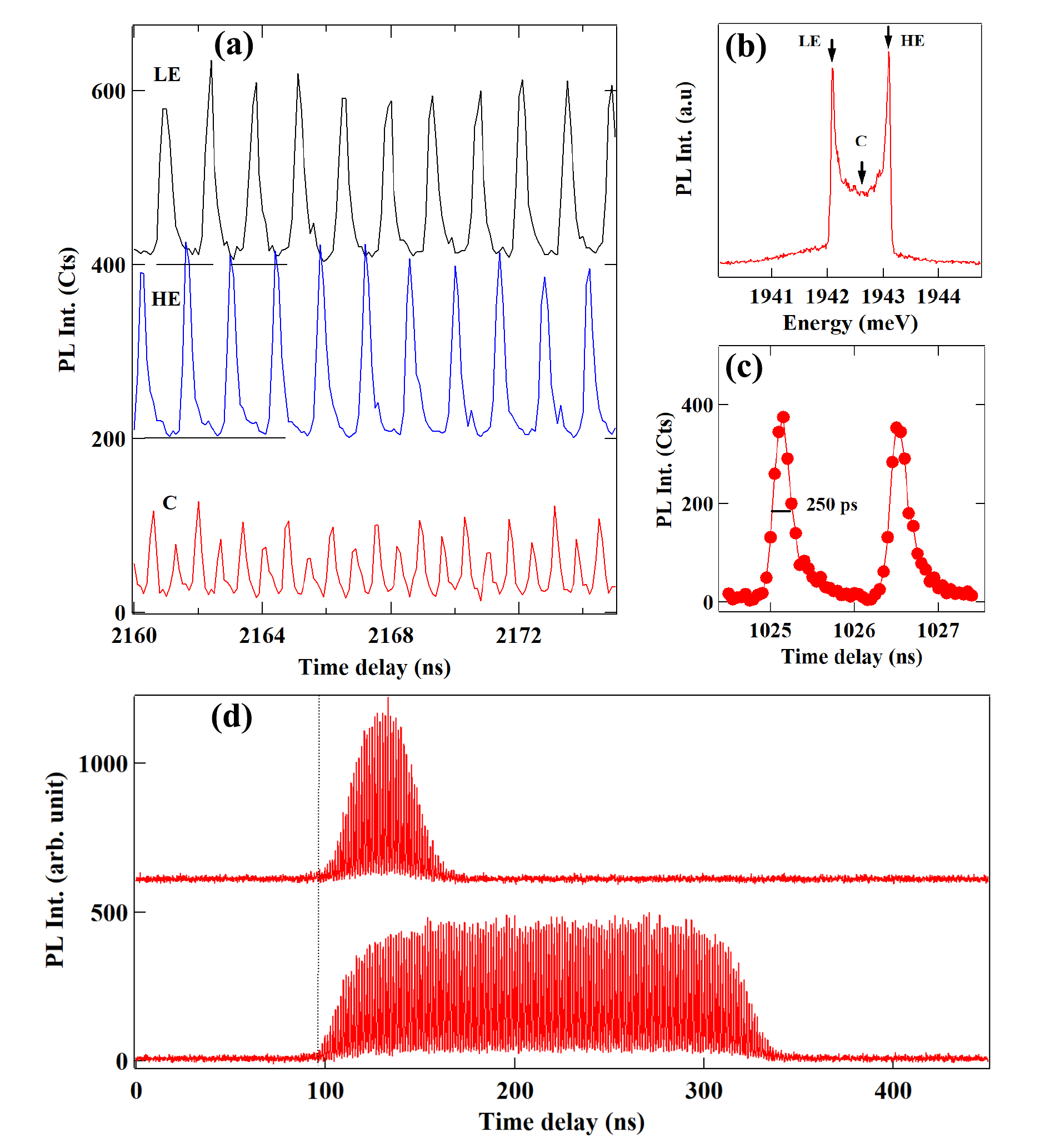}
\caption{Mechanical modulation induced by the strain field of a cw SAW at 0.718 GHz observed in the time domain on QD1. (a) Time evolution of the PL intensity recorded on the high energy side (HE), low energy side (LE) and center (C) of the mechanically broadened spectra presented in (b). (c) Detail of the time resolved modulation of the intensity of the HE side. (d) Observation in the time domain of the mechanical modulation induced by SAW pulses.}
\label{Fig5}
\end{figure}

\section{\label{sec4}Dynamic modulation of single quantum dots emission.}

With time-resolved photon counting phase-locked with the RF excitation, mechanical modulation of the PL of the QD by the SAW can be directly observed in the time domain. Figure \ref{Fig5}(a) shows the PL as a function of time for a $cw$ RF excitation at 0.718 GHz and for three different detection energies selected in the broadened peak of the time-integrated spectra (see Fig. \ref{Fig5}(b)). Oscillations of the intensity of the PL are observed for each detection energies as the SAW moves the transition energies of the QDs in and out of the considered detection range. The oscillations of the PL of the highest and lowest transitons (labelled HE and LE) are in opposition of phase. At the central detection energy (label C), the PL shows oscillations at twice the frequency of the SAW as the line crosses twice the detection window during each modulation cycle. 

The PL switching time, which is determined by the time spent by the PL line within the detection window, is around 250 ps for the experiment reported in Fig. \ref{Fig5}(c). Faster driving is certainly possible by increasing the SAW frequency. This illustrates the potential of SAW to be used for high frequency modulation of a single photon source \cite{Gell2008, Metcalfe2010}.

Under pulsed RF excitation, the 30 MHz frequency bandwidth of the IDT permits to generate mechanical pulses. As presented in Fig. \ref{Fig5}(d) the rise and fall time of the mechanical modulation observed in the PL of a dot is around 30 ns. Strain pulses with a full width at half maximum in the same time range can be generated (Fig. \ref{Fig5}(d)). 

\begin{figure}[hbt]
\centering
\includegraphics[width=3.3in]{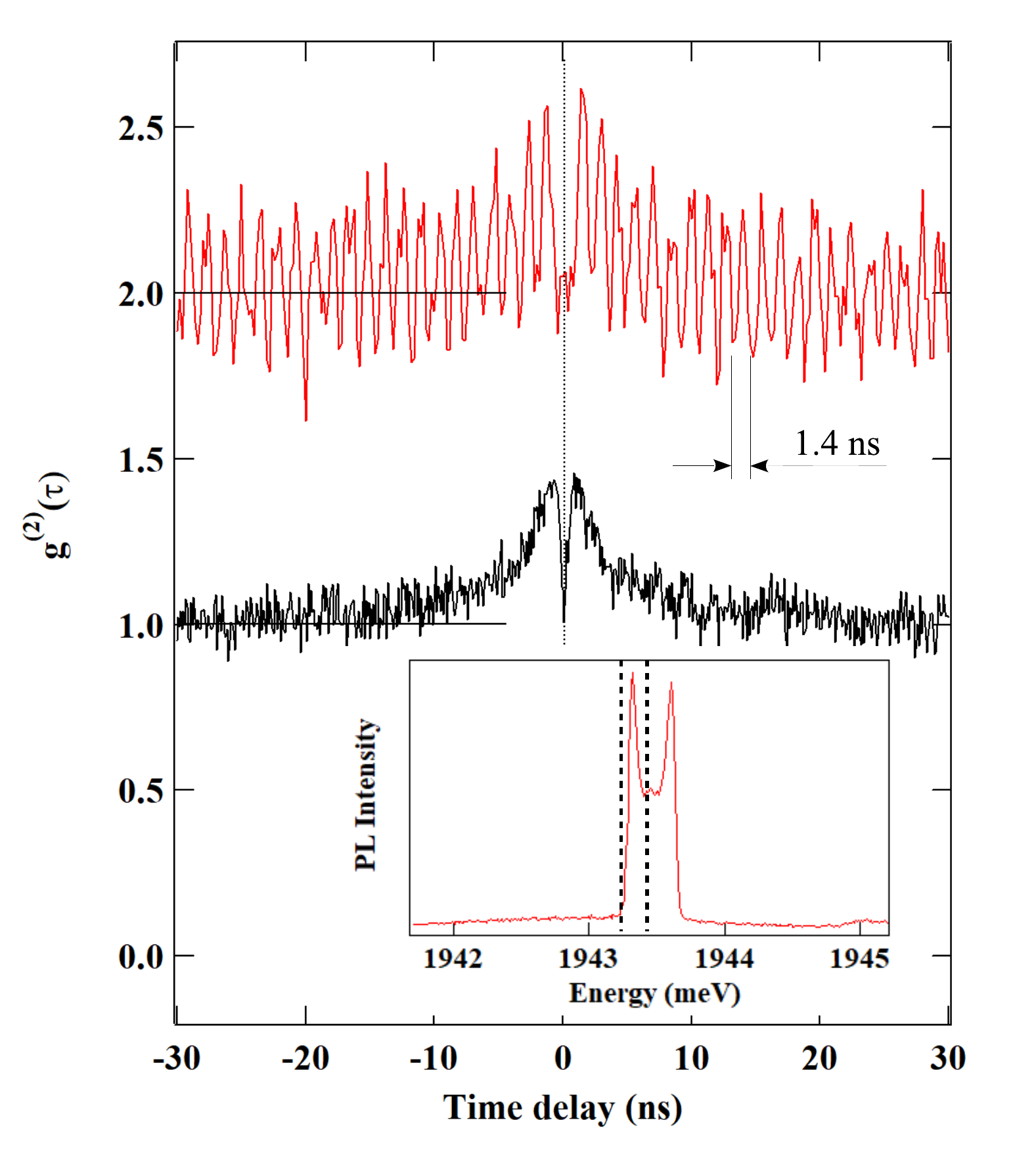}
\caption{Autocorrelation measurements of the PL intensity of a charged QD (QD1) without SAW excitation (black) and with a SAW modulation at 0.718 GHz and a RF power of 10 dBm (red). The red curve is shifted by 1 for clarity. The inset shows the energy detection window on the  mechanically broadened PL spectra.}
\label{Fig6}
\end{figure}

The time statistics of the photons emitted by a QD under mechanical modulation were analysed through autocorrelation measurements of the PL using a Hanbury Brown-Twiss set-up. The time resolution of the set-up is limited to about 700 ps by the combined jitter of the two APDs. Without SAW excitation, a dip is observed in the autocorrelation at zero delay. This antibunching is the signature of single photon emission from the QD. Its moderate depth is a consequence of the timing jitter of the detectors combined with the short lifetime of the excitons in these QDs which is around 250 ps. 

\begin{figure}[hbt]
\centering
\includegraphics[width=3.3in]{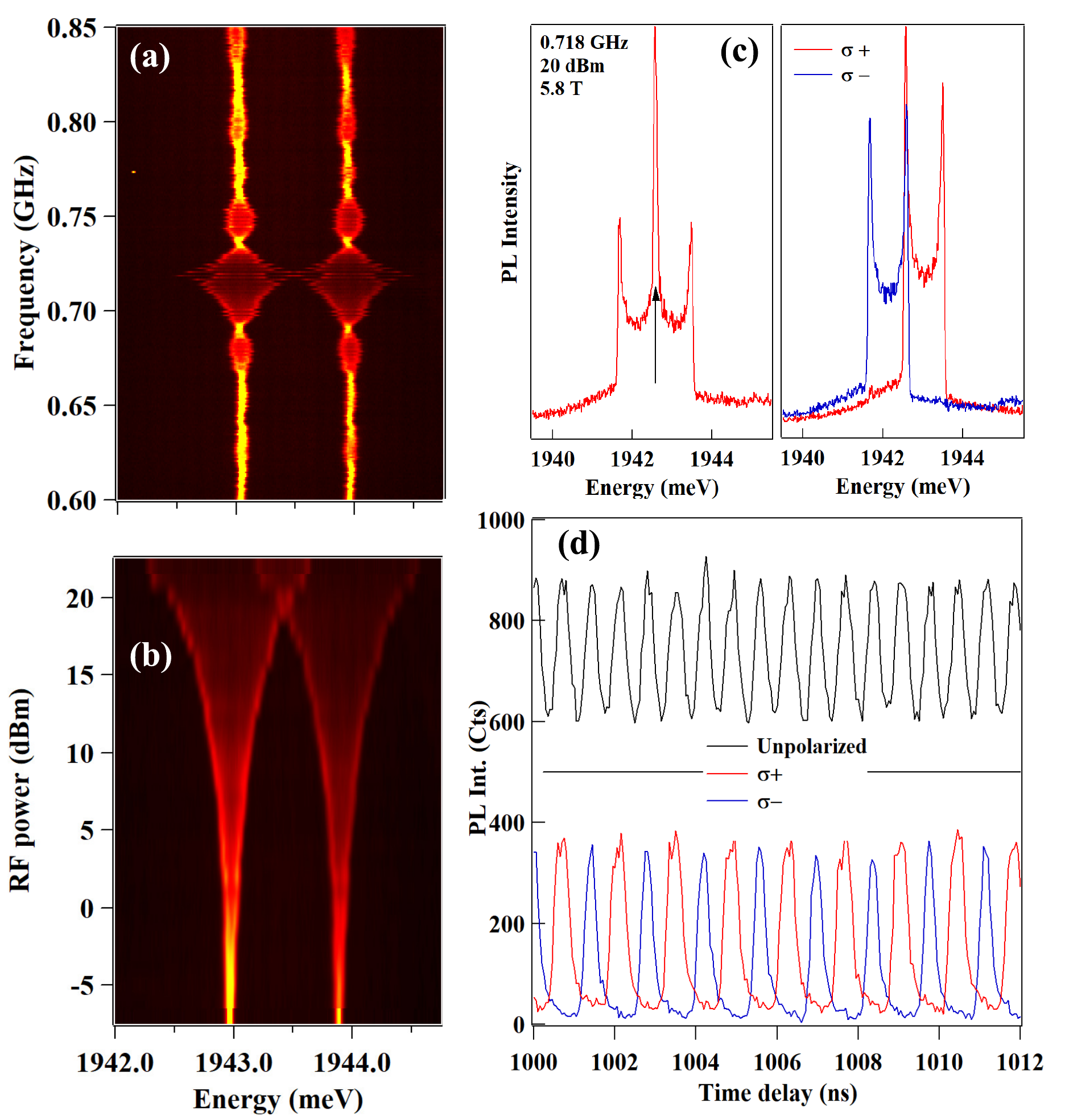}
\caption{(a) PL intensity map of a charged exciton (QD1) under a longitudinal magnetic field B$_z$=6T as a function of the energy and frequency of a RF excitation of 20dBm. (b) RF power dependence under B$_z$=6T obtained at a SAW frequency of 0.718 GHz. (c) PL spectra obtained at B$_z$=5.8T for a RF excitation power of 20dBm at 0.718 GHz. (d) Time evolution of the PL intensity of the central line (see (c)) for unpolarized and circularly polarized detection.}
\label{Fig7}
\end{figure}

The weak bunching observed at short delays with a full width of about 10 ns is related to  a blinking in the QD emission usually observed in charged QDs \cite{Sallen2010}. The  maximum  value of g$^{(2)}(\tau)$ increases  with  the  off/on  ratio  in  the  QD  emission and  the  decay  time  of  the  bunching  reveals  the  dynamics of charge fluctuations in the QD environment leading to the on/off behaviour.

Under $cw$ SAW excitation, autocorrelation measurements were done with the slits of the spectrometer adjusted to select the low energy side of the broadened emission spectra so that about 20 $\%$ of the emission was collected (see inset of Fig. \ref{Fig6}). In the autocorrelation signal we observe a dip at zero delay and clear oscillation with a period of about 1.4 nanoseconds. This oscillation corresponds to the signature of the opto-mechanical coupling observed in the photon counting regime \cite{Wieb2016}. It confirms that a single photon source can be efficiently modulated by a SAW excitation.

Under a magnetic field applied along the QD growth axis, the emission of the charged exciton is split by the Zeeman energy controlled by the Landé factor of the electron in the excited state and of the hole in the ground state. This leads to well resolved and perfectly circularly polarized emission lines. Both Zeeman split levels are modulated by the strain field of the SAW. At a given magnetic field and with a well chosen RF frequency and power, the Zeeman splitting can be compensated by the strain induced shift of the QD emission (Fig. \ref{Fig7}(a) and (b)). Corresponding emission spectra under a magnetic field B$_z$=5.8 T are presented in Fig. \ref{Fig7}(c).

When the Zeeman energy is compensated, most of the QD emission arises from a central peak (Fig. \ref{Fig7}(c)). The emission recorded on this central line is modulated at 1.436 GHz, twice the frequency of the SAW (black curve in Fig. \ref{Fig7}(d)). It corresponds to trains of antibunched photons (i.e. single photon source) with alternating circular polarization  modulated at the frequency of the SAW (blue and red curves in Fig. \ref{Fig7}(d)).

\section{\label{sec5}SAW superposition detected by single quantum dots.}

A QD acting as a nanometer-scale strain sensor can be used to monitor propagation and superposition of SAW pulses in an acoustic cavity. In our case the cavity is formed by two IDTs separated by 500 $\mu m$. The first IDT serves as the SAW source and the second IDT, when it is not connected to an external circuit, acts as a Bragg mirror for the propagating SAW and mechanical pulses are partially reflected back.

\begin{figure}[hbt]
\centering
\includegraphics[width=3.3in]{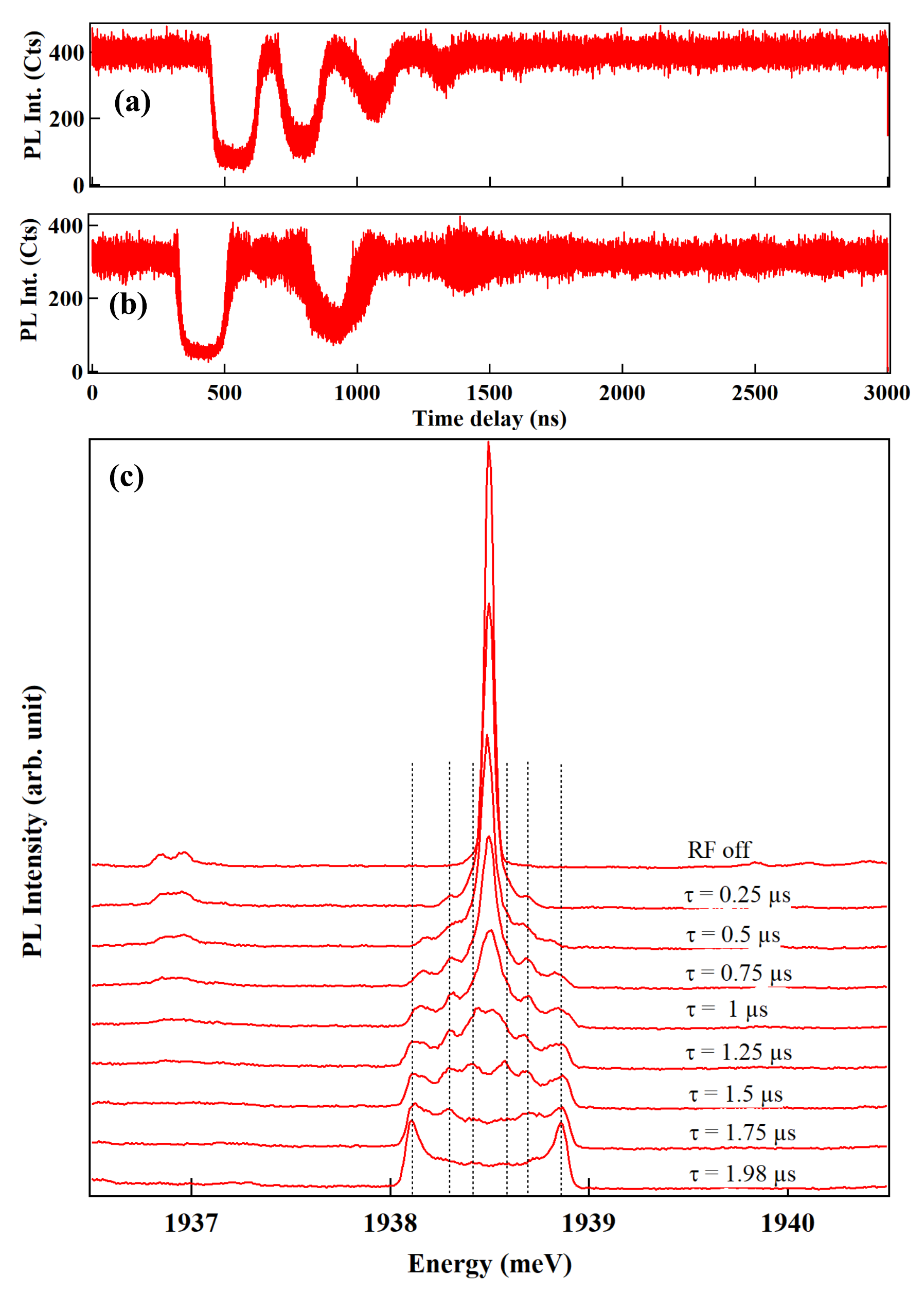}
\caption{Propagation and reflection of 175 ns SAW pulses observed (a) in the PL of a QD in the middle of the two IDTs (QD3) and (b) for a QD 5 $\mu m$ from IDT1 (QD1). (c) PL spectra of QD3 obtained under pulsed RF excitation with a repetition period of 2 $\mu s$ at a fixed excitation power (21 dBm) and frequency (0.713 GHz) and variable pulse duration.}
\label{Fig8}
\end{figure}

Figure \ref{Fig8} (a) and \ref{Fig8} (b) present the strain field probed in QDs at two different positions between the IDTs for a pulsed RF excitation of 175 ns at a repetition period of 3 $\mu s$. For a dot located in the middle of the two IDTs (Fig.\ref{Fig8}(a)), the propagation of the incident and up to three distinct reflected mechanical pulses are clearly observed in the intensity of the central line which decreases as the emission broadened (see Fig.\ref{Fig8}(c)). As expected, for a dot located a few $\mu m$ away from IDT1 (Fig.\ref{Fig8}(b)), the incident pulse arrives about 125 ns before (about one half of the travelling time between IDT1 and IDT2 $(1/2)l_{cav}/v_S\approx125ns$). In this configuration, the reflected acoustic pulse on IDT2 overlaps with its reflection on IDT1. Such overlap can lead to interferences of acoustic strain fields. 

\begin{figure}[hbt]
\centering
\includegraphics[width=3.3in]{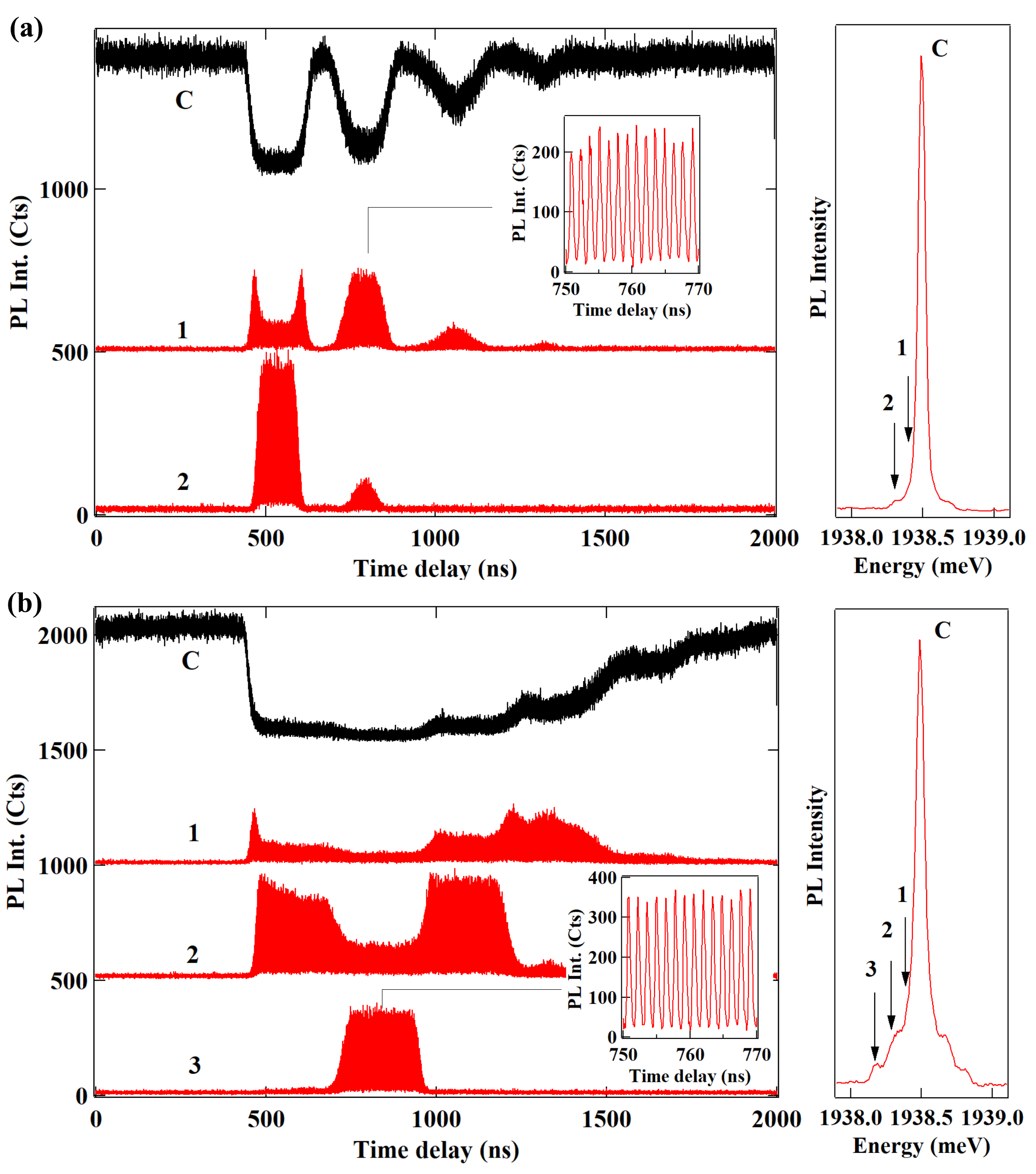}
\caption{Propagation and reflection of SAW pulses at 0.713 GHz observed on QD3 for two different lengths of pulses (a) 175 ns and (b) 500 ns and fixed RF power (21 dBm) and repetition period (2 $\mu s$). The mechanical modulation is detected in the central line (top black curve) and on the different side bands (see arrows in the corresponding time-integrated spectra). Insets show detail of the mechanical oscillation within the SAW pulses.}
\label{Fig8bis}
\end{figure}

Interferences in the acoustic cavity can be first observed in time-averaged spectra for a pulsed RF excitation with a fixed periodicity and variable pulses duration. As presented in Fig. \ref{Fig8}(c), time-averaged spectra present an increasing broadening with the increase of the RF excitation pulse length. At short pulse length, discrete values of 2$\Delta E$ are observed. At long RF pulses, close to the repetition period of 2$\mu s$, the characteristic broadened spectra observed under $cw$ excitation is recovered (bottom spectra in Fig. \ref{Fig8}(c)).

This discrete broadening arises from the overlap of pulses inside the acoustic cavity. For a dot in the middle of the cavity, when the length of excitation pulses becomes larger than $l_{cav}/v_S\approx$250ns, the incident and reflected mechanical pulses can overlap at the QD location and interfere. With constructive interference, the maximum amplitude of the strain field increases and produces an additional shift of the emission line larger than the one induced by the incident or reflected pulses alone.  

\begin{figure}[hbt]
\centering
\includegraphics[width=3.3in]{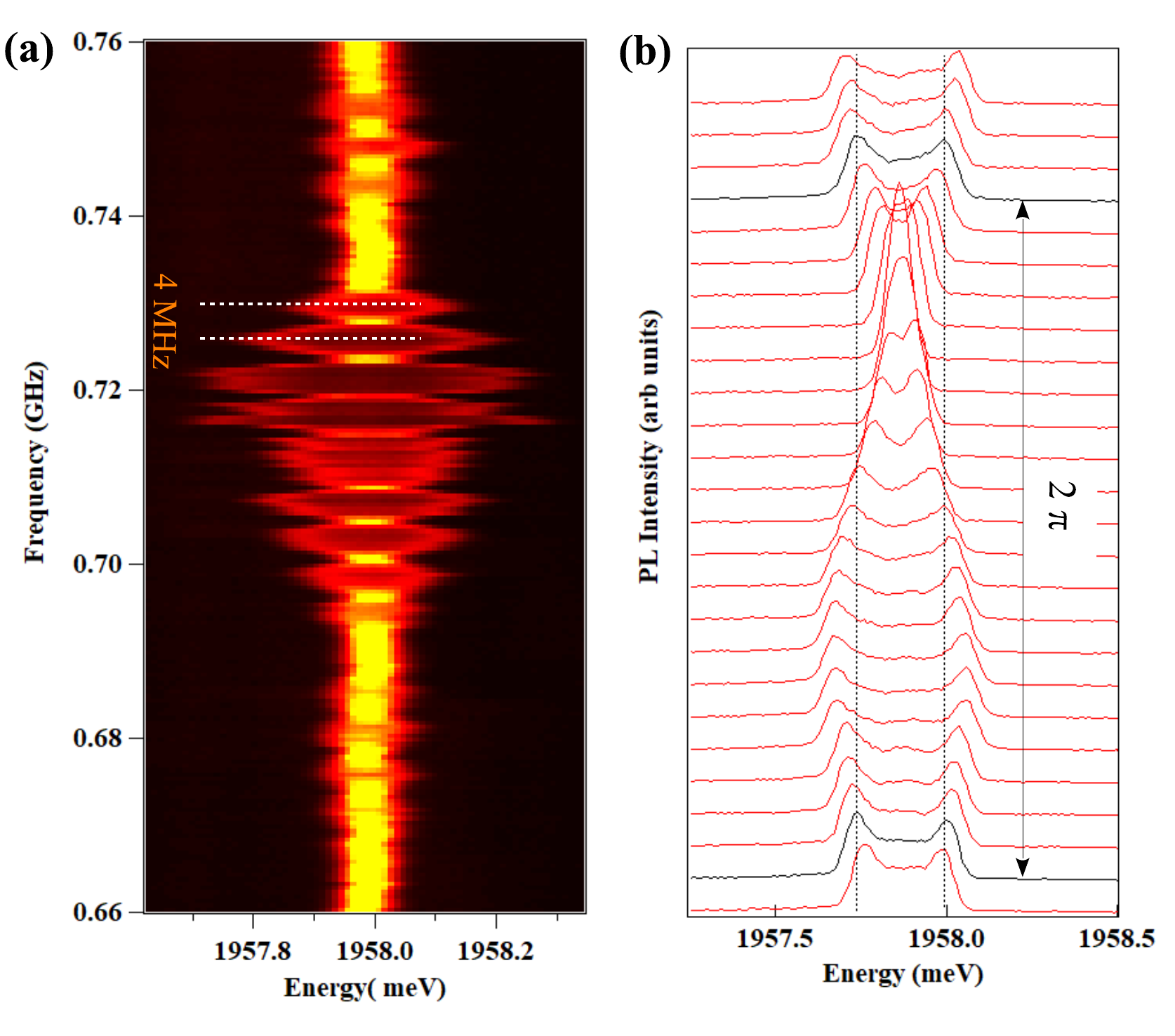}
\caption{Strain field superposition observed with counter-propagating SAWs generated by two IDTs excited with 9 dBm RF power. (a) Map of the PL intensity of a charged QD (QD4) as a function of the RF excitation frequency. (b) PL of QD4 strained by two counter-propagating SAWs at 0.723 GHz with identical amplitude as a function of the phase difference $\Delta \phi$ at the QD position.}
\label{Fig9}
\end{figure}

A detection of the mechanical modulation on the side-bands in the time domain permits to distinguish between the contribution of the different SAW pulses (incident and multi-reflected). This is illustrated in Fig. \ref{Fig8bis} for two lengths of pulses, 175 ns and 500 ns, and a fixed repetition rate of 2$\mu s$. For 175 ns pulses, incident and reflected pulses do not overlap. The largest energy shift (detection energy (2)) is induced by the direct incident pulse. Detecting at an intermediate energy (detection energy (1)) reveals the contribution of the first and second reflected pulses (red curves in Fig. \ref{Fig8bis}(a)). 

The situation is more complex for an excitation with 500 ns RF pulses (Fig. \ref{Fig8bis}(b)). The largest energy shift observed in time integrated spectra (detection energy (3)), which is larger than the one observed with 175 ns pulses (Fig. \ref{Fig8bis}(a)), arises from interferences between the end of the incident pulse and the beginning of the first reflected pulse. This is confirmed by the clear mechanical oscillation observed at detection energy (3) (see inset of Fig. \ref{Fig8bis}(b)) and the corresponding dip observed in the incident pulse when the mechanical modulation is detected at position (2). At longer delay, multi-reflected pulses are attenuated and the emission returns progressively to the central line (see black curve corresponding to a detection on (C) in Fig. \ref{Fig8bis}(b)).
 
The second IDT can also be connected to a RF power source and be used to generate SAWs. Figure \ref{Fig9} presents the broadening of the time-integrated emission spectra as a function of the RF excitation frequency obtained when the two IDT are excited with the same RF power (around 9 dBm). In the frequency range where the two IDTs produce almost the same strain field at the QD location, beats are clearly observed with a period of about 4 MHz. This period corresponds to a change of 2$\pi$ of the relative phase of the two counter-propagating sound waves at the QD location.

As illustrated in Fig. \ref{Fig9}(b) tunable interference between counter-propagating SAWs can also be achieved at a fixed RF excitation frequency and amplitude by tuning the relative phase of the RF excitation of the two IDTs \cite{Wieb2018}. A pattern similar to that of a standing wave is observed as $\Delta \phi$ is tuned with a RF delay line. For some value of $\Delta \phi$, destructive interference of the two SAWs occurs at the QD position and the emission remains unaffected by the strain. With an additional phase of $\pi$ interferences become constructive and the strain induced shift observed in time-average emission is maximum. This phase tuning provides an additional tool to control the strain field applied on an individual QD and, in particular, permits to tune the emission energy of the single photon source formed by a charged exciton.

\section{\label{sec6}Conclusion}

We have shown that high frequency SAW can be generated on ZnTe structures containing QDs. The emission energy of charged excitons in QDs is an efficient tool to probe the local strain field of the SAW. Strain in the 10$^{-4}$ range can be obtained on the QD layer. We observed in the time domain propagation of mechanical pulses in the 100 ns range in cavities and demonstrated modulation of the energy of individual CdTe/ZnTe QDs by phase controlled standing waves.

If required for particular applications, the SAW frequency could be further increased to a few GHz by reducing the size of the IDTs using the same fabrication technique. For instance, a pitch of 250 nm remaining much lager than typical grain size of ZnO could be reached for a mechanical modulation around 2.2 GHz. SAW excitation could be combined with resonant optical pumping techniques on magnetic QDs \cite{Tiwari2020} and the in-plane strain component $\epsilon_{11}$ of SAW pulses be used for the coherent mechanical driving of the spin state of an individual magnetic atom.

\begin{acknowledgements}{}

This work was realized in the framework of the Commissariat \`{a} l'Energie Atomique et aux Energies Alternatives (Institut de Recherche Interdisciplinaire de Grenoble) / Centre National de la Recherche Scientifique (Institut N\'{e}el) joint research team NanoPhysique et Semi-Conducteurs. The work was
supported by the French ANR project MechaSpin (ANR-17-CE24-0024). V.T. acknowledges support from EU Marie Curie grant
No 754303. The work in Tsukuba has been supported by the Grants-in-Aid for Scientific Research on Innovative Areas "Science of Hybrid Quantum
Systems" and for Challenging Exploratory Research.

\end{acknowledgements}

\section*{Data Availability Statement}

Data available on request from the authors.

\begin{appendix}

\section{Modelling of a SAW propagating along [110] in ZnTe.}

We present here the main results of an analytical model which permits the determination of the strain field of a SAW propagating along [110] in ZnTe. This model is used first for the dimensioning of the inter-digitated transducers and then to estimate the energy shift of the emission of a QD induced by the strain field of a propagating SAW.

\subsection{SAW velocity.}

For a SAW (Rayleigh wave) propagating along the [110] direction in a non-piezoelectric zincblende lattice, according to reference \cite{Schuetz2015} and \cite{BookOndesAcoustiques}, the displacement vector $\overrightarrow{u}$ in the basis ($x_1,x_2,x_3$), where $x_1$ is along [110] and $x_3$ is perpendicular to the surface (see Fig.~\ref{Fig1}(b)), is described by:

\begin{eqnarray}
\label{displacement}
u_{1}=A\left(e^{-qkx_3-i\varphi}+H.c.\right)e^{ik(x_1-v_St)}\\ \nonumber
u_{2}=0\\ \nonumber
iu_{3}=A\left(\gamma e^{-qkx_3-i\varphi}+H.c.\right)e^{ik(x_1-v_St)}
\end{eqnarray}

\noindent The adjustable parameter A controls the maximum amplitude of the displacement. $v_S$ is the velocity for propagation of SAW along [110] and $k=2\pi/\lambda_{S}$. $\lambda_{S}=v_ST_{S}$ is the wavelength of SAW with $T_{S}$ the period of the SAW. $v_S$ is given according to reference \cite{Schuetz2015} by

\begin{eqnarray}
v_S=\sqrt{\frac{X C_{11}}{\rho}}
\end{eqnarray}

\noindent and X is controlled by the equation

\begin{eqnarray}
(1-\frac{C_{11}}{C_{44}}X)(\frac{C_{11}C'_{11}-C_{12}^2}{C_{11}^2}-X)^2=\\ \nonumber
X^2(\frac{C'_{11}}{C_{11}}-X)
\end{eqnarray}

\noindent with $C'_{11}=(C_{11}+C_{12}+2C_{44})/2$. The $C_{ij}$ are the elastic coefficient of the material. X can be determined graphically. With the parameters of ZnTe listed in table \ref{table1}, we obtain:

\begin{itemize}
\item at T=300K:  $X\approx0.318$ and $v_S\approx1959 m.s^{-1}$
\item at T=0K:  $X\approx0.3165$ and  $v_S\approx1987 m.s^{-1}$
\end{itemize}

Rayleigh waves are a mixing of transverse and longitudinal phonons. The velocity of SAW is however slower than for transverse acoustic phonons ($v_{TA}\approx$2300 ms$^{-1}$ in ZnTe). As there is no matter above the surface, the average stiffness for the propagating surface phonon mode is reduced and the speed slows down. In our devices this speed is expected to be slightly affected by the presence of the thin ZnO layer which as a larger stiffness.

\begin{table}[htb] \centering
\caption{Parameters of ZnTe. For the elastic moduli C$_{ij}$ the first value is measured at T=300~K and the second extrapolated for T=0~K \cite{Landolt}.}
\label{table1}\renewcommand{\arraystretch}{1.0}
\begin{tabular}{p{1.5cm}p{1.5cm}p{1.5cm}p{1.5cm}p{1.5cm}}
\hline\hline
C$_{11}$         & C$_{12}$       & C$_{44}$       & $\rho$             &   e$_{14}$       \\
71.3/73.7 $GPa$  & 40.7/42.3 $GPa$& 31.2/32.1 $GPa$& 5908 $Kg/m^3$      &   0.025 $Cm^{-2}$ \\

\hline\hline
\end{tabular}
\end{table}

\subsection{Strain field of the SAW.}

The strain tensor $\epsilon$ is defined as

\begin{eqnarray}
\epsilon_{kl}=\frac{1}{2}\left(\frac{\partial u_k}{\partial x_l}+\frac{\partial u_l}{\partial x_k}\right)
\label{strain}
\end{eqnarray}

\noindent where $u_i$ is the displacement along the Cartesian coordinates $x_i$. For a SAW the displacement is given by equations \ref{displacement} where q, $\gamma$ and $\varphi$ can be determined as follows:

\begin{itemize}

\item  q satisfies the equation:

\begin{eqnarray}
(C'_{11}-\rho v_S^2-C_{44}q^2)(C_{44}-\rho v_S^2-C_{11}q^2)\\ \nonumber
+(C_{12}+C_{44})^2q^2=0
\end{eqnarray}

\item  $\gamma$ is given by:

\begin{eqnarray}
\gamma=q\frac{C_{12}+C_{44}}{C_{44}-C_{11}(X+q^2)}
\end{eqnarray}

\noindent with $X=\rho v_S^2/C_{11}$

\item  and $\varphi$ satisfies:

\begin{eqnarray}
e^{-2i\varphi}=-\frac{\gamma^\star-q^\star}{\gamma-q}
\end{eqnarray}

\end{itemize}

With the parameters of ZnTe we obtain $q\approx0.46+0.53i$, $\gamma\approx-0.63+01.19i$ and $\varphi\approx1.02$. 

Finally, the real components of the displacement (\ref{displacement}) are given by:

\begin{eqnarray}
u_{1}=2Ae^{(-gkx_3)}cos(hkx_3+\varphi)cos(kx_1-\omega t)\\ \nonumber
u_{2}=0\\ \nonumber
u_{3}=2A r e^{(-gkx_3)}cos(hkx_3+\varphi-\theta)sin(kx_1-\omega t)  \nonumber
\end{eqnarray}

\noindent where we defined $q=g+ih$, $\gamma=r e^{i\theta}$ and $\omega=2\pi/T_{S}$.

From these displacements calculated in the basis ($x_1,x_2,x_3$) we can deduce the strain in the same coordinate system using equation \ref{strain}. Results obtained with A=0.2 nm are presented in Fig. \ref{Fig1}.

\subsection{SAW induced energy shift of the exciton.}

The strain field produced by a propagating SAW modifies the energy of a confined exciton. This energy shift arises from a change of the heavy-hole exciton band gap. This shifts can be described by the Bir and Pikus Hamiltonian \cite{kp}.

For a strain field expressed in the in the cubic lattice basis (x=[100],y=[010],z=[001]), the shift of the conduction band is given by

\begin{eqnarray}
\Delta E_c=a_c(\epsilon_{xx}+\epsilon_{yy}+\epsilon_{zz})
\end{eqnarray}

\noindent where $a_c=-3.96eV$ is the conduction band deformation potential of CdTe. 

The shift of the band edge energy of the heavy hole is given by

\begin{eqnarray}
\Delta E_v=
-a_v(\epsilon_{xx}+\epsilon_{yy}+\epsilon_{zz})-\frac{b}{2}(\epsilon_{xx}+\epsilon_{yy}-2\epsilon_{zz})
\end{eqnarray}

\noindent with the valence band deformation potential of CdTe $a_v=0.55eV$ and $b=-1.0eV$ \cite{Adachi2005}. The valence band (conduction band) energies are defined to be positive for the downward (upward) direction of the energy. 

The change in the energy transition from the conduction to the heavy hole band  $\Delta E_{g,hh}$ is then given by 

\begin{eqnarray}
\nonumber
\Delta E_{g,hh}=\Delta E_c+\Delta E_v=\\
a(\epsilon_{xx}+\epsilon_{yy}+\epsilon_{zz})-\frac{b}{2}(\epsilon_{xx}+\epsilon_{yy}-2\epsilon_{zz})
\end{eqnarray}

\noindent with $a=a_c-a_v$. The strain in the cubic lattice basis (x,y,z) can be deduced from the strain in (x$_1$,x$_2$,x$_3$) by a rotation around z of -$\pi/4$.

For a rotation of an angle $\alpha$ around z, a tensor X' in R' is transformed in a tensor X in R by the rotation matrix $P_R^{R'}=(a'_{ij})$ according to

\begin{eqnarray}
X=P_R^{R'}X'
\end{eqnarray}

\noindent with

\begin{eqnarray}
P_R^{R'}=\left(
 \begin{array}{ccc}
\cos\alpha           &-\sin\alpha        &0            \\
\sin\alpha           &\cos\alpha         &0            \\
0                    &0                  &1            \\
\end{array}
\right) =(a'_{ij})
\end{eqnarray}

For the strain tensor $\epsilon$ (rank two tensor) we obtain

\begin{eqnarray}
\epsilon_{ij}=a'_{ik}a'_{jl}\epsilon'_{kl}
\end{eqnarray}

\noindent with the standard summation on the repeated indexes and $\epsilon$ in the crystal basis (x,y,z) reads:

\begin{eqnarray}
\epsilon_{xx}=1/2\epsilon_{11}\\ \nonumber
\epsilon_{yy}=1/2\epsilon_{11}\\ \nonumber
\epsilon_{zz}=\epsilon_{33}\\ \nonumber
\epsilon_{yz}=\sqrt2/2\epsilon_{13}\\ \nonumber
\epsilon_{xz}=\sqrt2/2\epsilon_{13}\\ \nonumber
\epsilon_{xy}=1/2\epsilon_{11}\\ \nonumber
\end{eqnarray}

For a SAW propagating along O$x_1$=[110] all the strain terms are different of zero and $\epsilon_{xx}-\epsilon_{yy}=0$. $\epsilon_{xy}$ can be used to control the Cr spin states $S_z=\pm1$ \cite{Vallin1974}.

With the strain tensor written in the basis (x$_1$,x$_2$,x$_3$), the energy shift becomes:

\begin{eqnarray}
\Delta E_{g,hh}=(a_c-a_v)(\epsilon_{11}+\epsilon_{33})-\frac{b}{2}(\epsilon_{11}-2\epsilon_{33})
\end{eqnarray}

Time dependent x$_3$ profile of the energy shift calculated with A=0.2 nm is presented in Fig. \ref{Fig4}.

\end{appendix}

\end{document}